 \documentclass[manuscript]{aastex63}
\received{\today}
\revised{}
\accepted{}
\submitjournal{ApJ}

\shortauthors{Sisti et al.}

\begin{document}

\title{Detecting Reconnection Events in Kinetic Vlasov Hybrid Simulations Using Clustering Techniques}

\correspondingauthor{Manuela Sisti}
\email{manuela.sisti@univ-amu.fr}

\author[0000-0003-3331-0606]{Manuela Sisti}
\affiliation{Aix-Marseille University, CNRS, PIIM UMR 7345, Marseille, France, EU}
\affiliation{Dipartimento di Fisica, Universit\`a di Pisa, Italy, EU}

\author{Francesco Finelli}
\affiliation{Dipartimento di Fisica, Universit\`a di Pisa, Italy, EU}

\author{Giorgio Pedrazzi}
\affiliation{HPC Department, Cineca, Bologna, Italy, EU}

\author{Matteo Faganello}
\affiliation{Aix-Marseille University, CNRS, PIIM UMR 7345, Marseille, France, EU}

\author{Francesco Califano}
\affiliation{Dipartimento di Fisica, Universit\`a di Pisa, Italy, EU}

\author{Francesca Delli Ponti}
\affiliation{HPC Department, Cineca, Bologna, Italy, EU}




\begin{abstract}

Kinetic turbulence in magnetized space plasmas has been extensively studied via {\it in situ} observations, numerical simulations and theoretical models. In this context, a key point  concerns the formation of coherent current structures and their disruption through magnetic reconnection.
We present automatic techniques aimed at detecting reconnection events in large data set of numerical simulations.
We make use of clustering techniques known as K-means and DBscan (usually referred in literature as unsupervised machine learning approaches), and other methods based on thresholds of standard reconnection proxies. All our techniques use also a threshold on the aspect ratio of the regions selected.
We test the performance of our algorithms. We propose an optimal aspect ratio to be used in the automated machine learning algorithm: AR=18.
The performance of the unsupervised approach results to be strongly competitive with respect to those of other methods based on thresholds of standard reconnection proxies.

\end{abstract}



\section{Introduction}
Magnetic reconnection is an ubiquitous phenomenon observed 
in the laboratory as well as in many space and astrophysical environments. It is the energy driver for solar flares and coronal mass ejections \citep{Priest_1982}.  It occurs routinely at the boundary between the solar wind and the Earth’s magnetosphere \citep{dungey1961interplanetary, PriestForbes_2000}, both at dayside and at magnetotail, or at the flanks induced by the presence of large-scale Kelvin-Helmholtz vortices \citep{faganello2017}. An important outcome is the injection of accelerated particles into the magnetosphere. In the context of turbulent plasmas, such as solar wind \citep{haynes2014reconnection,osman2014} and magnetosheat plasma \citep{retino2007}, reconnection plays a central role representing a possible alternative path, as compared to the usual vortex-vortex interaction, for energy transfer towards smaller and smaller scales \citep{Cerri2017, karimabadi2013coherent, camporeale2018coherent}. It is behind many of the risks associated with space weather such as, for example, disturbs to Global Navigation Satellite System signals, electronic damage to satellites or power grids \citep{cassak2016}. 

Magnetic reconnection is probably the most known and important process in magnetized plasmas for which Magnetohydrodynamics (MHD) accurately describe the evolution at large scales, where the magnetic field is frozen in the fluid motion of the plasma and its topology is preserved. In this context, magnetic reconnection is the only process being able to rearrange the magnetic field topology on a global, MHD scale despite it occurs in a very narrow region of typical size of the order of the ion kinetic scale length. The magnetic rearrangement, being on the large MHD scale, is accompanied by a strong energy release due to the conversion of magnetic energy into plasma flow, particle acceleration, heating, and wave generation.
The classical picture of reconnection is obtained by considering a 2D domain where a large scale magnetic field, the so called equilibrium, presents an inversion line where it changes its direction. This is a 1D equilibrium configuration, since $\partial_x=\partial_z=0$ while the inversion takes place along the $y$ direction, that corresponds to a kinetic equilibrium for the distribution functions known as the Harris sheet \citep{harris1962}. The 1D magnetic field can be represented as ${\bf B} = B_{_x}(y) \, {\bf e}_x$, directed along the $x$-axis and varying along the $y$-axis being zero along the neutral line at a given $y = y_{_0}$. The corresponding {\em out-of-plane} current has a peak and the region around is dubbed {\em current sheet} (hereafter CS) or {\em current layer}. In certain favorable condition this layer becomes unstable to the 2D reconnection instability and the ideal MHD laws are locally violated \citep{Furth1963, White_1980, PriestForbes_2000} leading to the breaking and reconnection of field lines and eventually to the formation of an X-point like structures characterized by an inflow/outflow ion and electron fluid velocity advecting the magnetic flux toward/away from the reconnecting regions. Often in the laboratory but also in space, a mean {\em out-of-plane} magnetic field, ${\bf B}_{_{gf}} = B_{_0} \, {\bf e}_z$, is associated to the CS. 
In this case the dynamics is quasi-2D and reconnecting lines do not ``cut\&past'' but actually slip across the CS changing their original connectivity, even if their projection still seems to break and re-connect. This configuration, the so-called guide-field regime, is  often studied by means of 2D simulations.

In such a system, magnetic reconnection can be easily recognized. On the contrary, as soon as CSs have more complex shapes, to ascertain the presence of reconnection is far less simple, e.g when CSs are dynamically generated by 2D \citep{henri2012,Daughton2014} or 3D \citep{faganello2014,borgogno2015,sisti2019satellite} Kelvin-Helmholtz vortices or, even worse, by magnetized turbulence \citep{Servidio2010, Zhdankin2013}. For each of these cases a different {\it ad hoc} method has been developed, based on a deep knowledge of the peculiarity of each system.

For investigating the complex non-linear dynamics of such systems, numerical simulations are widely used, providing also an important support for the understanding of satellite data. However this approach generates an impressive  amount of data to deal with, in particular when working with kinetic simulations spanning the entire phase space. Large data sets, as well as the difficulty of easily recognize reconnecting structures in these sets, suggest a possible application of machine learning for reconnection analysis. Recently, in \cite{Dupuis_2020}, some machine learning techniques based on signatures in the particle velocity distribution function in the phase space have been tested to identify reconnection regions. Moreover, in \cite{Hu_2020} for the same purpose to automatically find magnetic reconnection sites, supervised machine learning methods based on CNN (Convolutional Neural Network) are used. In \cite{Hu_2020} it has been shown that the supervised machine learning approach turns out to perform very good and promising and that the algorithm seems to be capable of detecting reconnection even in cases in which the human labelling has failed. However, the set up of a large labelled training dataset is a long-term  process which requires the contribution of a number of experts in the field. On the other hand, an unsupervised algorithm doesn’t require a training dataset thus being more flexible and quicker to be implemented  (but not simpler). Here we analyse the performances of unsupervised algorithms in order to prove that such an approach can be an efficient method to automatically detect reconnection. 
For this purpose, we set up an hybrid 2D-3V (two-dimensional in space, three-dimensional in velocity space) simulation with kinetic ions and fluid electrons, describing the turbulent dynamics of a collisionless plasma where magnetic structures and CSs, continuously generated by plasma motions, interact together and eventually coalesce or disrupt due to magnetic reconnection. Beyond the physical interest of this simulation set on typical solar wind turbulence parameters, it represents a good test for a machine learning approach since reconnection occurs in a variety of different configurations far from the idealized 1D Harris-like.

Aiming at developing an automatic procedure for the detection of physical structures of basic interest, such as CSs and reconnecting structures, we have developed the following techniques. The first group relies on ``standard'' unsupervised machine learning techniques, such as K-means (in particular Lloyd's algorithm) \citep{MacQueen1967} and DBscan (Density Based Spatial Clustering of Applications with Noise) \citep{Ester1996}. The second group instead is based on a number of standard proxies used in the literature, and the definition of the corresponding thresholds, as markers to detect and highlight the presence of a reconnection event. In particular we cite: current density, electron vorticity and decoupling of the electron dynamics from the magnetic field. 
Note that all methods use physical quantities particularly suited to be considered as a signature of reconnection. These quantities, extracted here from numerical simulations, are the ones usually measured by on-board instruments of in-situ satellites. All methods are based on a final fundamental step for selecting reconnecting structures: the definition of a threhsold on the aspect ratio of the CSs. This threshold is motivated by the physics of the reconnecting CSs whose shape (their typical lenght and width) is not random but imposed by the development of the reconnection instability. 

There exist several models of reconnection that could be used to extrapolate a reasonable value for this threshold. The simplest one is the resistive Sweet-Parker model \citep{Sweet1958,Parker1957} where the CS length $L$ and width $\delta$ depend on the local reconnection rate $R$, 
i.e. on the outflow velocity \citep{cassak_liu_shay_2017}. Actually, the reconnection rate predicted from the Sweet-Parker model is too slow to account for observations and simulations in collisionless plasmas. Other models such as the so-called ``fast reconnection'' predict a reconnection rate of the order of $\sim 0.1$, quite in agreement with observations and simulation results recently found in the literature \citep{cassak_liu_shay_2017}. 
A different approach for the estimation of the CS aspect ratio relies on the role of the tearing mode as a sufficient condition for reconnection to occur. By looking at the wavenumber $k_m$ of the most unstable mode in a CS of width $\delta$, $k_m$$\delta \sim 2\pi \delta / L \sim 0.5$,  we get $\delta / L \sim 0.08$ \citep{karimabadi2005}, not  far from $\sim 0.1$. 
In other words, 
 there is a direct link between the aspect ratio and the reconnection rate. Therefore, we make use of the aspect ratio to distinguish structures where most probably reconnection is on-going  using it as a threshold. 

The paper is organized as follows. In Section 2  the Hybrid Vlasov-Maxwell model (HVM) and the 2D-3V simulation data are introduced. In Section 3 we describe the methods developed to automatically identify CSs and reconnecting regions. In particular in Section 3.1 we discuss about our main method which uses the unsupervised machine learning techniques. In Section 3.2 and 3.3 we present other two alternative methods which don't use machine learning but are still automatic algorithms to locate reconnecting regions using standard reconnection proxies. 
In Section 4 we discuss the performances of these different methods in finding reconnection sites.  Finally, our results are summarized in Section 5.

\section{Simulation}
\label{sec:simulation}

To identify a wide panorama of reconnection sites emerging dynamically and not prepared ad-hoc, we make use of a 2D-3V plasma turbulence simulation and use an hybrid model with kinetic ions and fluid electrons (with mass) \citep{valentini2007}. This simulation is run using the HVM code based on an Eulerian approach to solve the ion Vlasov equation \citep{MangeneyJCP2002} coupled to the Maxwell equations but neglecting the displacement current and adopting the quasi-neutral approximation, $n_i \simeq n_e \simeq n$.
We use ion quantities to normalize the equations. These are: the ion mass $m_i$, the ion cyclotron frequency $\Omega_{ci}= e {\bar B}/m_i c$, Alfv\'en velocity $v_A = {\bar B}/ \sqrt{4 \pi m_i n }$ and so 
the ion skin depth $d_i = v_A\Omega_{ci}$. The electron skin depth reads $d_e = \sqrt{m_e/m_i}$, where $m_e$ is the electron mass. Finally ${\bar n}$, ${\bar B}$, ${\bar E}$ and ${\bar f}$ are the normalizing density, magnetic field, electric field and distribution function, respectively. Then, the Vlasov equation for the ion distribution function $f = f(x,y, v_x, v_y, v_z,t)$ reads:

\begin{equation}{\label{eq:vlasov}}
\frac{\partial f}{\partial t} + {\bf v} \cdot \frac{\partial f}{\partial {\bf x}} + ({\bf E} + {\bf v} \times {\bf B})\cdot\frac{\partial f}{\partial {\bf v}}=0
\end{equation}

The Ohm's equation for the electron response reads:

\begin{equation}{\label{eq:ohm}}
\footnotesize
{\bf E} - d_e^2 \nabla^2 {\bf E} = - {\bf u_e} \times {\bf B} - \frac{1}{n} \nabla P_e + 
d_e^2 \nabla \cdot\left[n\left({\bf uu}-{\bf u}_e{\bf u}_e\right)\right] 
\end{equation}

where ${\bf u} = \int{f{\bf v} \, d{\bf v}} \, / \int{f d{\bf v}}$ and ${\bf u}_e$ are the ion and the electron fluid velocities, respectively. Furthermore $n$ is given by $n = \int{f d{\bf v}}$. Finally the dimensionless Faraday and Ampere laws read:

\begin{equation}{\label{eq:faraday_ampere}}
\footnotesize
\nabla \times {\bf B} = {\bf J} \,; \;\;\; \frac{\partial {\bf B}}{\partial t} = - \nabla \times {\bf E}
\end{equation}

 
For the sake of simplicity, we take an isothermal closure: $P_e = nT_{0e}$. In this approach electron inertia (terms proportional to $d_e^2$ in Eq. \ref{eq:ohm}) is a key ingredient since it allows for reconnection to occur by decoupling the electrons from the magnetic field.

\begin{figure}
   \centering
   \includegraphics[width=\textwidth]{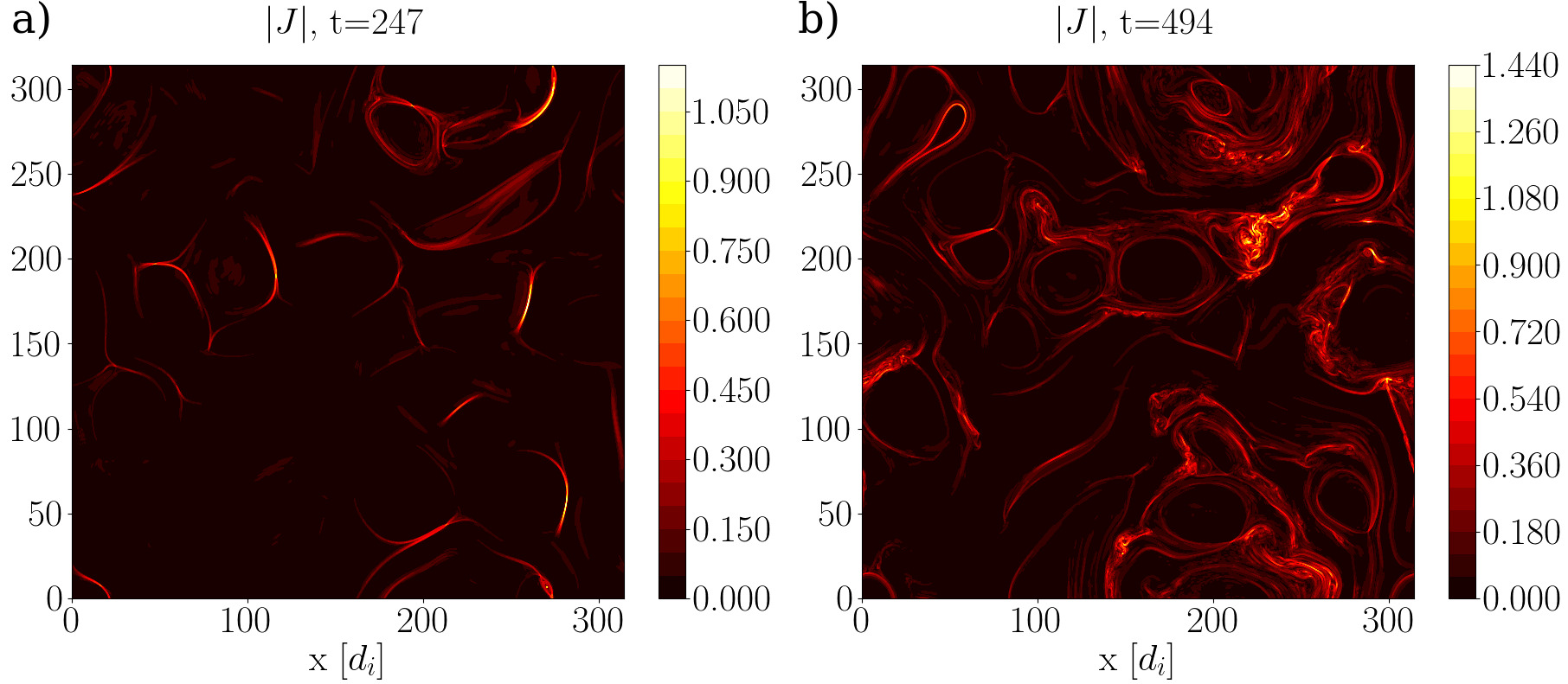}
      \caption{Snapshot of the contour plots of current density $|{\bf J}|$  for two different times of our simulation: 
      a)  $t=247$ $[1/\Omega_{ci}]$ (current sheets formed), b) $t=494$ $[1/\Omega_{ci}]$ (fully-developed turbulence).}
         \label{figure:current}
   \end{figure}

We take a squared simulation box with $L = L_x = L_y = 2 \pi\, 50$ using $N_x = N_y = 3072$ points. The corresponding physics goes from the MHD fluid-like behavior of the largest wavelength $\lambda \sim L$ to the ion kinetic physics $\lambda \sim d_i$ (= 1 in dimensionless units) to the electron inertial scale, $\lambda \sim d_e$. In terms of the corresponding frequencies we are across the ion cyclotron physics (= 1 in dimensionless units). We set the magnitude of the initial guide field along the $z$-direction equal to one. The initial ion distribution function is given by a Maxwellian with corresponding uniform temperature. The electron temperature is set equal to that of the ions, $T_{0i} = T_{0e}$. As a result, the plasma beta, the ratio between the fluid to the magnetic pressure, is equal to one (still in dimensionless units). We sample the velocity space using $51^3$ uniformly distributed grid point spanning $[-5v_{th,i}, 5v_{th,i}]$ in each direction, where $v_{th,i} = \sqrt{\beta_{i}/2}$ is the initial ion thermal velocity. The mass ratio is $m_{\rm i}/m_{\rm e} = 100$ allowing us to well separate the ion ($d_{i}$) and electron ($d_{e}$) skin depth. The simulation is initialized by adding a isotropic magnetic perturbation given by the sum of sinusoidal modes with random phase.  The corresponding wave number interval is $k\in[0.02,0.12]$. The corresponding root mean square value of the magnetic field is $\delta B_{rms} \simeq 0.28$. In the initial phase, the initial perturbation starts to produce a number of non interacting CSs where, later, at about one eddy turnover time $\sim 250$, associated to the largest perturbed wavelength, reconnection become effective. After about two ion turn over times the system reaches a fully developed turbulent state. Two snapshots of the current density are shown in Fig.\ref{figure:current} corresponding to the different stages: when CSs are well formed (left frame) and start to reconnect, and the fully developed turbulent state (right frame).

\section{Methods} \label{sec:methods}
In order to speed up the identification of CSs and reconnecting structures, we develop three different techniques. The main one uses unsupervised machine learning, and will be addressed hereafter as ``AML''. Two other alternative algorithms, which don't use machine learning, are developed in order compare the performances; they will be addressed as ``A1'' and ``A2''. A comparison between these three methods will be presented hereafter. The physical quantities we use as variables for detecting reconnection in AML are the current density magnitude $J=|{\bf J}|$, the magnitude of the in-plane electron fluid velocity $|{\bf u}_{e,\mathrm{in-plane}}|$, the magnitude of the in-plane magnetic field $|{\bf B}_\mathrm{in-plane}|$, the magnitude of the electron vorticity $\Omega_e=|{\bf \Omega}_e|=|\nabla\times{\bf u}_e|$, 
the electron decoupling term $E_{\mathrm{dec}, e}=|({\bf E} + {\bf u}_e \times {\bf B})_z|$ and the energy conversion term ${\bf J}\cdot ({\bf E} + {\bf u}_e \times {\bf B})$.  The first three variables are  related to the geometry of the CSs. The electron decoupling term describes the de-freezing of magnetic field lines from the electron fluid motion. The last variable is a proxy which accounts for the energy dissipation at the reconnection sites \citep{zenitani2011}. In A1 and A2 we use only some of them, in particular in A1 $J$, while in A2 $J$, $\Omega_e$ and $E_{dec,e}$. 
 
The aspect ratio (hereafter AR) of reconnecting CSs is not random but it's linked to reconnection development, thus all our algorithms have a common step: a threshold on the AR of the CSs selected. We define the AR as 

     \begin{equation}\label{eq:AR}
     AR = \frac{\text{length}}{\text{width}}
 \end{equation}
 for each structure found. We give an estimation for the CSs width and length using the automated method explained in \cite{Califano_2020} and references therein. 
 Note that for methods A1 and A2 the CSs are defined as region where the current density $J$ is greater than a certain thresholds, as defined in \cite{Zhdankin2013}, while
 in AML, we make use of a different technique to define the CSs in the physical space (as explained in the following sections).
 Anyway, once a CS is defined as a collection of grid points in the 2D physical space, the automated procedure computes the Hessian matrix $H$ of $|{\bf J}|$ at the current peak (the local maximum of $J$). We look at the interpolated profile of $J$ along the direction given by the eigenvector associated to the largest eigenvalue of $H$ and define the CS's width as the full width at half maximum of $J$. A similar procedure for computing the length (i.e. interpolating $J$ along the direction of the eigenvector associated to the smallest eigenvalue) would be misleading because in a turbulent system CSs are seldom ``straight'' along that direction. We give an automated estimation for the length by computing the maximal distance between two points belonging to the same structure, similarly to what done in \citet{Zhdankin2013} where, however, the CSs are defined using a threshold on the current density.

It is worth noticing that all the variables we use are available in satellite data sets. In particular $|{\bf u}_{e}|$ and $|{\bf B}|$ are directly measured by on-board instruments, while $J=|n({\bf u} - {\bf u}_e)|$, $E_{\mathrm{dec}, e}=|({\bf E} + {\bf u}_e \times {\bf B})_z|$ and ${\bf J}\cdot ({\bf E} + {\bf u}_e \times {\bf B})$ are simple algebraic combinations of measured quantities. 
$|{\bf u}_{e,\mathrm{in-plane}}|$ and $|{\bf B}_\mathrm{in-plane}|$ can be computed by setting the out-of-plane direction aligned with the average (local) magnetic field. 
Finally $\Omega_e$ and AR can be calculated by measured quantities and spacecrafts' relative positions in the case of multi-satellite missions as CLUSTER or MMS \citep{dunlop2002,fadanelli2019}.

\subsection{AML}
\begin{figure}
   \centering
   \includegraphics[width=0.6\textwidth]{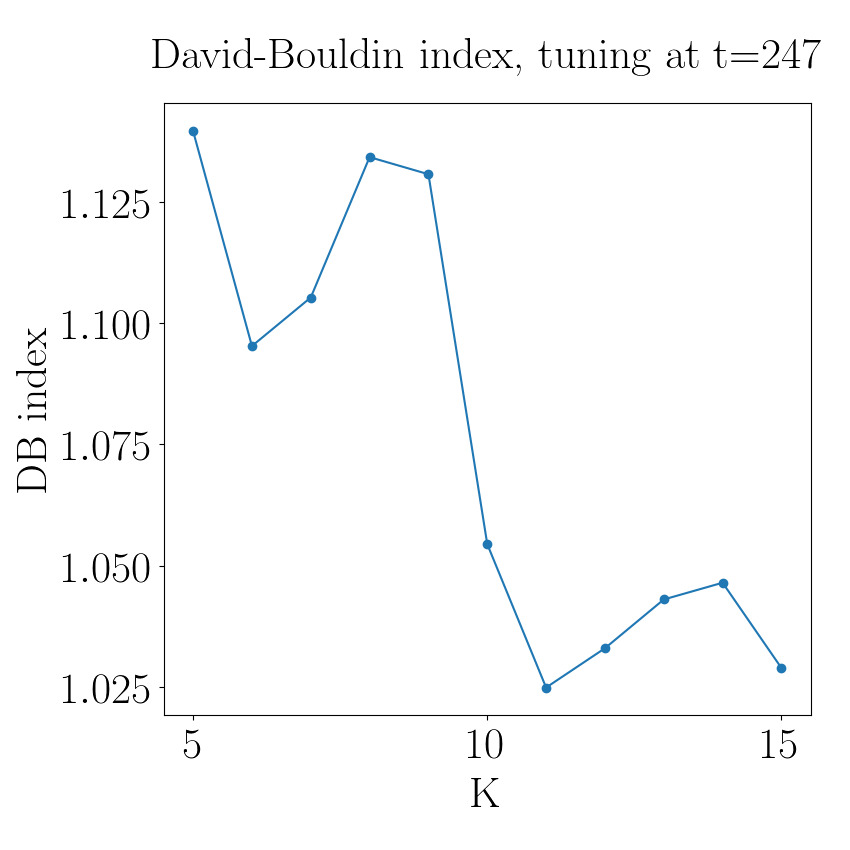}
      \caption{Value of the David-Bouldin index as a function of possible ``K'' from $K=5$ to $K=15$, for $t=247$ $[1/\Omega_{ci}]$. The David-Bouldin index has a minimum for $K=11$.
      }
         \label{figure:tuning}
   \end{figure}

The core of the first method is constituted by two unsupervised machine learning algorithms: K-means (Lloyd's algorithm) and DBscan (Density Based Spatial Clustering of Applications with Noise). These are both clustering techniques aimed at learning a grouping structure in a data-set. K-means algorithm requires to set the parameter ``K'' corresponding to the number of clusters we expect to find in our data-set in the variable space. To fix the parameter ``K'' we make a pre-processing step in which we tune the best value of ``K'' for our data. Our approach then summarizes in the following steps: 
    
    1) Pre-processing step for the tuning of the ``K'' parameter for the K-means algorithm using a cross-validation-like approach   with the Davis-Bouldin index applied to the predicted clusters as internal cluster metric \citep{rhys2020machine}. The tuning  is applied in the variable space defined at the beginning of Section 3.
    We choose to tune the ``K'' value at about one eddy-turnover time ($t\simeq 247$ $[1/\Omega_{ci}]$), corresponding to the phase when CSs are well formed but still not interacting one each other. The result of the tuning is shown in Figure \ref{figure:tuning}. The best ``K'' value is the one for which the David-Bouldin index reaches its minimum, in our case it turns out to be $K=11$, as shown in Fig.\ref{figure:tuning}.
    
    2) K-means (Llyod's algorithm) is applied to the chosen variables which we list here again: the current density magnitude $J=|{\bf J}|$, the magnitude of the in-plane electron fluid velocity $|{\bf V}_{e,\mathrm{in-plane}}|$, the magnitude of the in-plane magnetic field $|{\bf B}_\mathrm{in-plane}|$, the magnitude of the electron vorticity $\Omega_e=|{\bf \Omega_e}|=|\nabla \times {\bf V}_e|$, the electron decoupling term (i.e., $E_{\mathrm{dec}, e}=|({\bf E} + {\bf V}_e \times {\bf B})_z|)$ and the energy conversion term ${\bf J}\cdot ({\bf E} + {\bf u}_e \times {\bf B})$. These variables are normalized between 0 and 1. This is a common requirement for many machine learning estimators, since in non-normalized data-set the presence of outliers could alter the result. The K-means algorithm returns $K=11$ clusters in the variable space, as we show in Table \ref{table:1} for time $t\sim 247$ $[1/\Omega_{ci}]$. 
    
    \begin{table}[h!]
    \footnotesize
    \centering
    \begin{tabular}{c |c c c c c c |r} 
    
     Cluster & $\overline{|{\bf J}|}$& $\overline{|{\bf V}_e|}$ &	$\overline{|{\bf \Omega}_e|}$	& $\overline{E_{dec,e}}10^{-2}$ & $\overline{|{\bf B}_{in-plane}|}$ & $\overline{|{\bf J}\cdot({\bf E} + {\bf V}_e \times {\bf B})|}10^{-3}$ &	Grid point number \\
    \hline

{\bf 1} & {\bf 0.369} & {\bf 0.253} & {\bf 1.239} & {\bf 0.031} & {\bf 0.138} & {\bf 0.930} & {\bf 37\,776}\\
2 & 0.059 & 0.125 & 0.162 & 0.021 & 0.402 & 0.170 & 160\,928\\
3 & 0.036 & 0.239 & 0.049 & 0.022 & 0.109 & 0.079 & 593\,803\\
4 & 0.033 & 0.173 & 0.055 & 0.021 & 0.242 & 0.060 & 660\,889\\
5 & 0.027 & 0.078 & 0.045 & 0.021 & 0.286 & 0.053 & 685\,840\\
6 & 0.031 & 0.145 & 0.040 & 0.021 & 0.049 & 0.062 & 813\,751\\
7 & 0.025 & 0.153 & 0.031 & 0.021 & 0.168 & 0.041 & 1\,165\,723\\
8 & 0.022 & 0.063 & 0.023 & 0.021 & 0.072 & 0.039 & 1\,243\,427\\
9 & 0.022 & 0.069 & 0.029 & 0.021 & 0.207 & 0.039 & 1\,278\,303\\
10 & 0.022 & 0.070 & 0.024 & 0.021 & 0.144 & 0.037 & 1\,325\,111\\
11 & 0.021 & 0.128 & 0.022 & 0.021 & 0.108 & 0.034 & 1\,471\,633\\

    \hline
    \end{tabular}
    \caption{The eleven clusters in the variable space of our data-set which result from the application of the K-means algorithm to $t\sim 247$ $[1/\Omega_{ci}]$. For each cluster, identified by an index, we report the mean value of our variables and the number of grid points which belong to it. 
    In bold the ``interesting'' cluster with the highest value of mean current density.}
    \label{table:1}
    \end{table}
    
\begin{figure}
   \centering
   \includegraphics[width=0.7\textwidth]{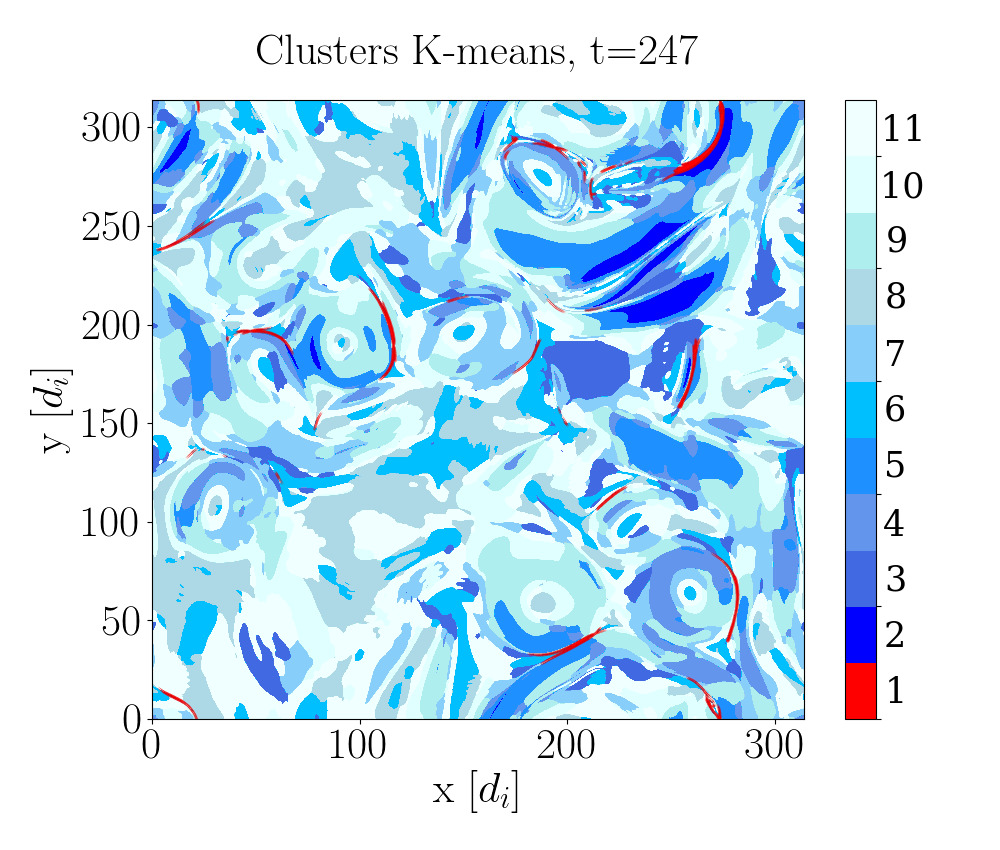}
   \includegraphics[width=0.7\textwidth]{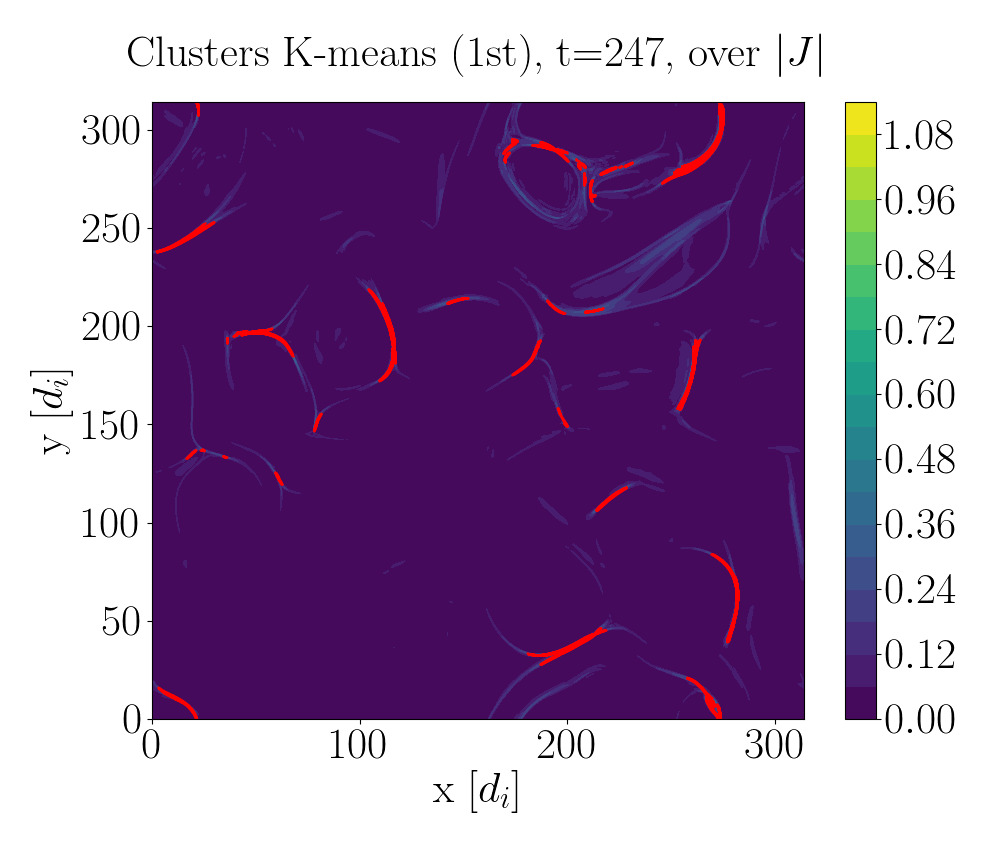}
      \caption{Top panel: the regions which constitute all eleven clusters in the variable space; in red cluster ``1'', while in various shades of blue all other clusters (please note that any of the colors refers to a physical variable). Bottom panel: the regions which constitute cluster ``1'' (the interesting one) in the variable space, in red, superimposed to the contour plot of current density $|{\bf J}|$. }
         \label{figure:kmeans}
   \end{figure}

    In Table \ref{table:1} we report for each cluster the mean value of our variables (in dimensionless units) and the number of grid points which belong to it.
    Among these clusters we choose to analyze the one with the highest mean value of the current density (since a necessary condition for reconnection to occur is the presence of a peak in the current density value), which is cluster ``1''. This particular cluster in the variables space is made-up of different structures in the physical space of our box. 
    In Figure \ref{figure:kmeans}, top panel, we draw in the simulation ($x,y$) space domain the shaded iso-contours of the eleven clusters calculated by K-means (in the variables space). Cluster 1 is represented by the red color, the others by the different blue variations. In the same Figure, bottom panel, we draw the shaded iso-contours of cluster 1 regions (red) superimposed to the contour shaded plots of the current density $|{\bf J}|$, suggesting that cluster 1 roughly corresponds to the ensemble of CSs.

It is worth noticing that other unsupervised methods are suitable for the same purpose such as, for instance, the K-medoids technique. In particular, by applying K-medoids at t=247 we did not observed any significant improvement of the results. Therefore, K-means turns out to be a very good approach in this context, a baseline for this kind of analysis.

\begin{figure}
   \centering
   \includegraphics[width=0.7\textwidth]{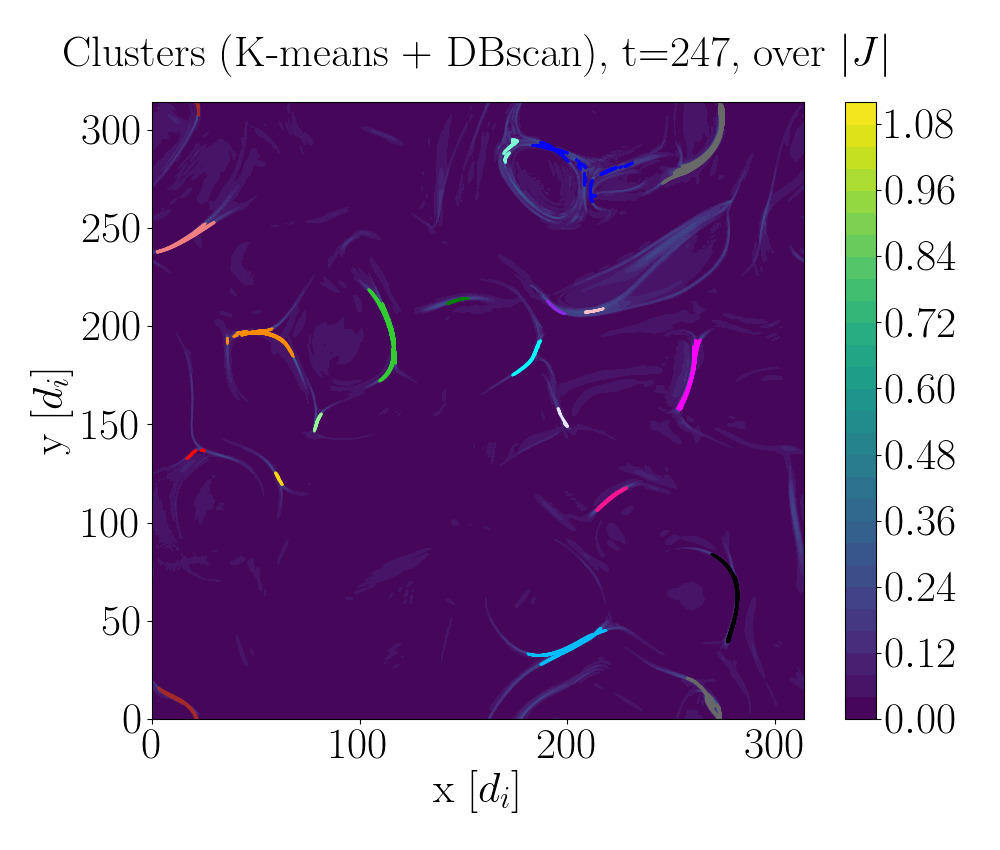}
      \caption{Contour plot of current density $|{\bf J}|$. In different colors the different structures identified through DBscan algorithm applied to cluster ``1'' of Table \ref{table:1}.}
         \label{figure:dbscan}
   \end{figure}

   3) DBscan. From Figure \ref{figure:kmeans} we see that cluster 1 in the variable space is composed by different structures in  physical space. Thus we need another algorithm able to distinguish different structures using their location in physical space. A technique which is suited for this scope is DBscan, which takes as input the x and y-coordinates of the points  belonging to cluster 1. We take as searching radius for DBscan $\epsilon=50$ grid points which correspond to about $5 d_i$, and $Min\_pts = 100$ points, as the minimum number of points for a single structure, where $\epsilon$ and $Min\_pts$ are intrinsic variables of the DBscan algorithm.
   The results obtained by using DBscan are shown in Figure \ref{figure:dbscan} where the different colors represent the different structures that are identified. 
   
   We underline the fact that DBscan is not used here for image analysis but only for ``cutting'' cluster 1 in different subsets corresponding, in physical space, to different structures. Figure 3 and 4 help humans in visualizing the correspondence between these subsets and CSs where reconnection could occur.
   
   4) Threshold on the AR of structures found. For each structure found at step 3 we compute the AR (see Section \ref{sec:methods}, Eq. \ref{eq:AR}). We consider different possible thresholds for the AR to get the better performance. In particular we have tried the following values: 10, 12.5, 20, 30, 50 and 70.
   
\subsection{A1}
A1 is the simplest algorithm to identify structures. It is based on two step:

1) A threshold on the current density $|{\bf J}|$ defined as $J_{\rm thr} = \sqrt{<J^2> + 3\sigma}$, where $\sigma = \sqrt{<J^4> - (<J^2>)^2}$, see \cite{Zhdankin2013}. More precisely, we select all regions in the physical space where the current density overcomes the threshold $J_{\rm thr}$. However, since a region of enhanced current is a necessary condition but not a sufficient one for reconnection, we add a second  step.

2) A threshold on the AR of the structures in order to select the most probably reconnecting structures. As done in AML, we consider different possible threshold values on AR: 10, 12.5, 20, 30, 50 and 70. 

\subsection{A2}
A2 is a refinement of A1. Another step is added in order to increase the precision in finding reconnection events. The method is summarized as follows.

1) As in A1 we fix a threshold on the current density and select the regions which overcome this threshold. 

2) We look at each structure and we select only the ones which include some points (at least two) which overcome both a threshold over electron vorticity ${\bf \Omega}_{e}$ and over $E_{dec,e}$. In particular these two thresholds are defined as: $\Omega_{e,\rm thr} = \sqrt{<\Omega_e^2> + 3\sigma_{\Omega}}$, where  $\sigma_{\Omega} = \sqrt{<\Omega_e^4> - (<\Omega_e^2>)^2  }$, and $E_{dec,e,\rm thr} = \sqrt{<E_{dec,e}^2> + 3\sigma_{E}}$, where  $\sigma_{E} = \sqrt{<E_{dec,e}^4> - (<E_{dec,e}^2>)^2  }$. 

3) Finally, as in A1, we set a threshold on the AR of the selected structures. Also in this case, we consider different possible values for the AR: 10, 12.5, 20, 30, 50 and 70.

\begin{figure}
    \centering
    \includegraphics[width=\textwidth]{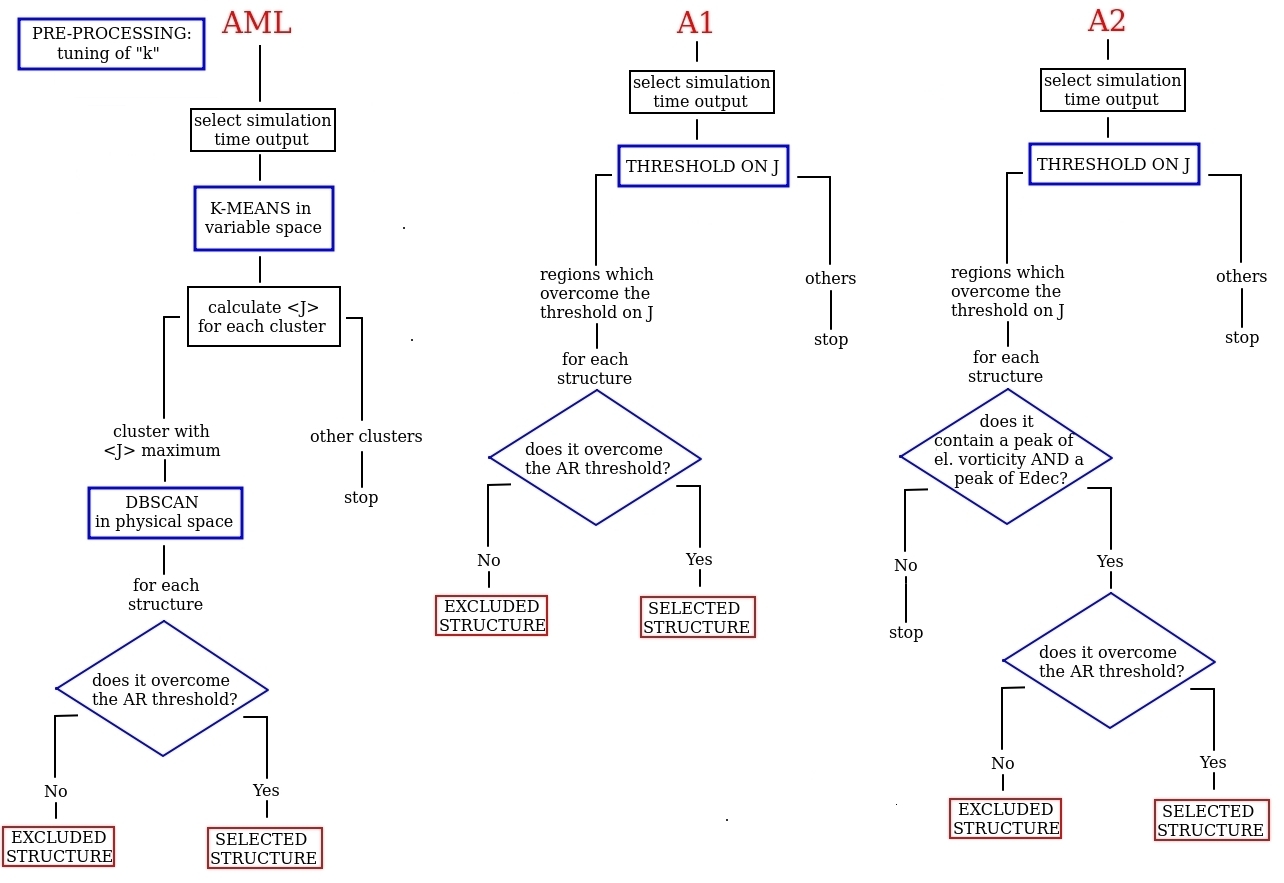}
    \caption{Flow-chart for the three methods we are considering.}
    \label{fig:flow-chart}
\end{figure}

\medskip
In Table \ref{table:methods_summary} we give a summary of the steps of each method, while in Figure \ref{fig:flow-chart} we show a flow-chart to sketch the selection procedure to detect possible reconnecting structures.

 \begin{table}
  \centering
  \begin{tabular}{c|l}
    Method & Steps\\
    \hline
    AML & 0) Selection of physical variables we want to use\\ 
      & 1) pre-processing: tuning of ``K'' value for the K-means\\
      & 2) K-means on variable space\\
      & 3) DBscan on physical space\\
      & 4) threshold on aspect ratio\\
      \hline
    A1 & 1) threshold on the current density $|{\bf J}|$ in the physical space\\
      & 2) threshold on the aspect ratio\\
      \hline
    A2 & 1) threshold on the current density $|{\bf J}|$ in the physical space\\
      & 2) request of points which overcome thresholds on $|{\bf \Omega}_e|$ and $E_{dec,e}$\\
      & 3) threshold on the aspect ratio\\
      \hline
  \end{tabular}
  \caption{Summary of the steps for each method.}
  \label{table:methods_summary}
\end{table}

\section{Results}

Once three sets of candidate reconnecting structures have been selected using the three different methods (see flow-chart in Figure \ref{fig:flow-chart}), we estimate the accuracy of these techniques by looking at candidate sites one by one. In particular, we check if reconnection is going on in each single site by looking if typical signatures of reconnection are present: inversion of the in-plane magnetic field, X-point configuration of the magnetic flux $\Psi$, converging electron inflows toward the X-point and diverging outflows, magnetic fluctuation along the guide-field direction, peaked electron decoupling and energy conversion terms.

Now, to compare quantitatively the performances of the three algorithms AML, A1 and A2 for different values of the AR threshold. We define the following quality-parameters: 

\begin{equation}
\text{Precision} = \frac{\text{\# reconnecting structures among selected structures}}{\text{ \# selected structures}}
\end{equation}
\begin{equation} 
\text{nMR-precision} = \frac{\text{\# non-reconnecting structures among excluded structures}}{\text{\# excluded structures}}
\end{equation}
where the selected/excluded structures are all those possible reconnection regions whose ARs overcome/don't overcome the AR threshold as schematized in the flow-chart in Figure \ref{fig:flow-chart}. 

The best algorithm performance is obtained  when both precision and nMR-precision (which stands for non-magnetic-reconnection precision) are $\sim 1$.  Indeed, in this case we would have at the same time that the algorithm has selected only reconnection sites and has not excluded reconnection sites.

\begin{figure}
   \centering
   \includegraphics[width=\textwidth]{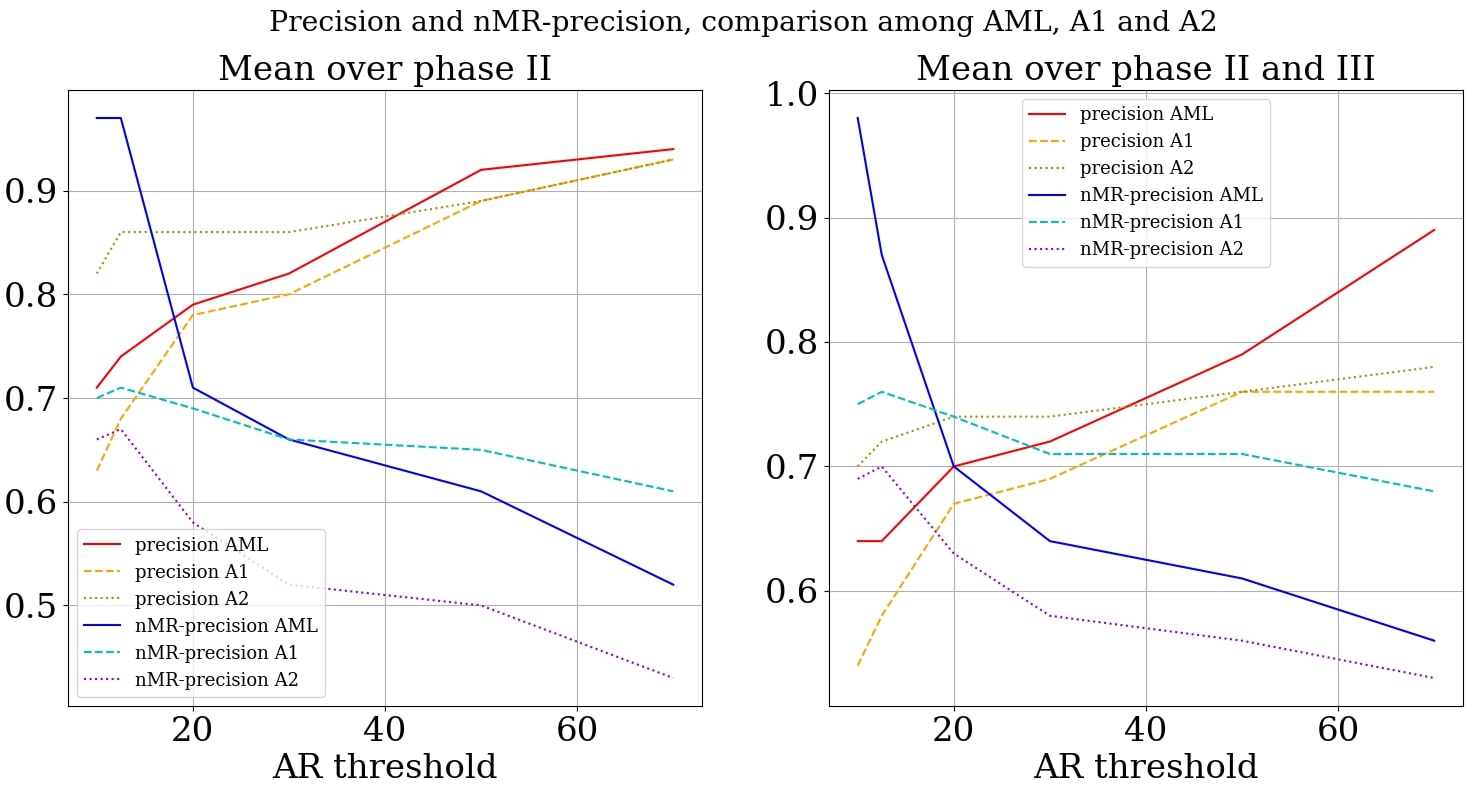}
      \caption{Plots of the values of precision (red, orange and brown lines) and nMR-precision (blue, cyan and violet lines) averaged over the three central times (left panel) and over four times (right panel), as a function of the AR threshold, for the three different Methods we are analyzing: solid line, AML; dashed line, A1; dotted line, A2.}
         \label{figure:comparison}
   \end{figure}

In Table \ref{table:method_1_ml}, \ref{table:method_2} and \ref{table:method_3} we show the values of these quality-parameters for AML, A1 and A2, respectively, and for different AR thresholds. The results are shown at five time instants of our simulation:  $t_1 \sim 20$ $[1/\Omega_{ci}]$ (initial phase, no evidence of  CS structures, $t_2\sim 230$ $[1/\Omega_{ci}]$, $t_3 \sim 247$ $[1/\Omega_{ci}]$, $t_4 \sim 282$ $[1/\Omega_{ci}]$ (CS developed regime), $t_5 \sim 494$ $[1/\Omega_{ci}]$ (fully developed turbulence). We will refer to these three distinct phases as phase I, II and III.
In the last two columns we list the mean value of our quality-parameters during phase II, $t = t_2$, $t_3$, $t_4$ and during phase II and III, $t = t_2$, $t_3$, $t_4$, $t_5$. Looking at Table \ref{table:method_1_ml} corresponding to AML we observe: a) precision increases when AR threshold increases, b) nMR-precision decreases when AR threshold increases, meaning that we are loosing ``good'' (reconnecting) sites, c) precision worsen when we include in our analysis phase III (turbulent regime). Same conclusions are drawn by looking at Table \ref{table:method_2} and \ref{table:method_3}, corresponding to methods A1 and A2. In order to ease the comparison for the different performances, we plot in Figure \ref{figure:comparison} precision (solid red, dashed orange and dotted brown lines, AML, A1 and A2, respectively) and nMR-precision (solid blue, dashed cyan and dotted violet lines, AML, A1 and A2, respectively) averaged over phase II (left panel) and over phase II and III (right panel) as a function of the AR threshold. By comparing these plots, left panel, we see that the unsupervised machine learning algorithm AML turns out to be strongly competitive with respect to the methods based on thresholds on standard proxies. In particular precision performs better for AML rather than for A1. For what concerns A2, we see that the precision appears to be less influenced by the
AR threshold variations. In particular, the precision of A2 (brown dotted line) turns out to be better at small AR thresholds than for AML; on the other hand, the nMR-precision (dotted violet line) worsen, so that lots of ``good'' reconnection sites are excluded. The same holds when including in the average the values at $t = t_5$ (fully developed turbulent phase), see the right panel of Figure \ref{figure:comparison}. However, and this is a quite general result, precision and nMR-precision worsen a bit when considering the full turbulent phase. This worsening is the consequence of the dynamics driven by the turbulence which advects, shrinks, breaks and deforms the CSs. This dynamics together with the merging of nearby structures typical of a 2D geometry makes much more difficult a correct estimation of the typical width and length. Moreover, turbulence also creates regions with sharp variations in current density where, however, reconnection is not occurring thus ``confusing'' the various Methods and decreasing their accuracy. Except for this worsening in precision at the end of the simulations, all Methods perform well both during the initial phase I when, correctly, no reconnection sites are detected, as well as during the central phase II when the CSs are well detected. According to us, it is possible to define an optimal AR threshold for AML and A1, which corresponds to a compromise between having a good precision and not loosing lots of good reconnection sites. For AML, this optimal AR threshold corresponds to the one at the intersection between the plot of precision (left panel of Figure \ref{figure:comparison}, red solid line) and the plot of nMR-precision (left panel of Figure \ref{figure:comparison}, blue solid line). In this case, we get AR $\simeq 18$, which gives a precision and nMR-precision accuracy of the order of $78\%$. For A1, the same is obtained for $AR \simeq 14$ with a score of $\simeq 71\%$. In both cases AR values are in agreement with theoretical estimations. For A2, it is not possible to make the same estimation of the optimal AR since the two plots do not intersect; this is a limitation of A2. Finally, in our simulation the performance of AML method improves a bit (reaching $80\%$) if we use, in the variable set, ${\bf J}\cdot ({\bf E} + {\bf u}_e \times {\bf B} + \frac{\nabla{P_e}}{ne})$ instead of ${\bf J}\cdot ({\bf E} + {\bf u}_e \times {\bf B})$. This can be due to the fact that in our model we consider a scalar pressure, thus, in the absence of off-diagonal terms in the pressure tensor, this term doesn't contribute to the effective Ohmic heating \citep{Birn_Hesse_2010}. 

\begin{table}[h!]
    \centering
    \begin{tabular}{c c|c c c c c | c c} 
    
     Tempo [1/$\Omega_{ci}$]& & 20 & 230 & 247 & 282 & 494 & Mean (3t) & Mean (4t)\\

    \hline
     N. structures & & 35 & 29 & 19 & 24 & 32\\
     \hline
     N. structures &$AR>10$  & 0 & 17 & 17 &  21 & 28 \\
     &$AR>12.5$ & 0 &  16 & 17 & 20 & 25 \\
     & $AR>20$ & 0 & 14 & 15 & 14 &  19\\
     & $AR>30$ & 0 & 10 & 10 & 13 &  12\\
     & $AR>50$ & 0 & 6 & 9 & 12 & 7\\
     & $AR>70$ & 0 & 3 & 7 & 6 & 4\\
     \hline
     precision & $AR>10$ & -- & 0.65 & 0.82 & 0.67 &  0.43 & 0.71 & 0.64\\ 
     &$AR>12.5$ & -- & 0.69 & 0.82 & 0.7 & 0.36 & 0.74 & 0.64\\
     & $AR>20$ & -- &  0.79 & 0.80 & 0.79 & 0.42 & 0.79 & 0.7\\
     & $AR>30$ & -- &  0.8 & 1 & 0.77 & 0.33 & 0.82 & 0.72 \\
     & $AR>50$ & -- & 1 & 1 & 0.75 & 0.43 & 0.92 & 0.79\\
     & $AR>70$ & -- & 1 & 1 & 0.83 & 0.75 & 0.94 & 0.89\\
    \hline
     nMR-precision& $AR<10$ & 1 & 0.92 & 1 & 1 &  1 & 0.97 & 0.98\\
     & $AR<12.5$ & 1 & 0.92 & 1 & 1 & 0.57 & 0.97 & 0.87\\
     & $AR<20$ & 1 & 0.93 & 0.5 & 0.7 & 0.69 & 0.71 & 0.7 \\
     & $AR<30$ & 1 & 0.79 & 0.55 & 0.64 & 0.6 & 0.66 & 0.64\\
     & $AR<50$ & 1 & 0.74 & 0.5 & 0.58 & 0.64 & 0.61 & 0.61\\
     & $AR<70$ & 1 & 0.65 & 0.42 & 0.5 & 0.68 & 0.52 & 0.56\\
     \hline
    
    \end{tabular}
    \caption{Number of structures found, precision and nMR-precision for AML, for different AR threshold (from 10 to 70). The results are shown for five different times of our simulation:  $t\sim20$ $[1/\Omega_{ci}]$, $t\sim 230$ $[1/\Omega_{ci}]$, $t\sim 247$ $[1/\Omega_{ci}]$, $t\sim 282$ $[1/\Omega_{ci}]$, $t\sim 494$ $[1/\Omega_{ci}]$. In the last two columns we report the mean value of our quality-parameters for the three central times (230, 247, 282) and for four times (230, 247, 282, and 494). In the second row we list the number of structures found at the end of the third step; in rows from 3rd to 8th we give the number of detected structures overcoming the specified AR threshold. In rows from 9th to 14th we give the values of precision for different AR threshold computed among the structures enumerated in rows 3-8; finally in rows 15th to 20th the same for nMR-precision.}
    \label{table:method_1_ml}
    \end{table}

    
    \begin{table}[h!]
    \centering
    \begin{tabular}{c c|c c c c c | c c} 
    
     Tempo [1/$\Omega_{ci}$]& & 20 & 230 & 247 & 282 & 494 & Mean (3t) & Mean (4t)\\
    \hline
     N. structures & & 11 & 45 & 47 & 52 & 454 &\\
     \hline
     N. structures &$AR>10$ & 0 & 22 & 23 & 18 & 88&\\
     &$AR>12.5$ & 0 & 20 & 20 & 17 & 70&\\
     & $AR>20$ & 0 & 15 & 15 & 12 & 43&\\
     & $AR>30$ & 0 & 13 & 12 & 10 & 26&\\
     & $AR>50$ & 0 & 11 & 9 & 7 & 11\\
     & $AR>70$ & 0 & 6 & 6 & 5 & 4\\
     \hline
     precision & $AR>10$ & -- & 0.68 & 0.61 & 0.61 & 0.26 & 0.63 & 0.54 \\
     & $AR>12.5$ & -- & 0.75 & 0.7 & 0.59 & 0.27 & 0.68 & 0.58\\
     & $AR>20$ & -- & 0.8 & 0.8 & 0.75 & 0.33 & 0.78 & 0.67\\
     & $AR>30$ & -- & 0.77 & 0.83 &  0.8 & 0.35 & 0.8 & 0.69\\
     & $AR>50$ & -- & 0.82 & 1 & 0.86 & 0.36 & 0.89 & 0.76  \\
     & $AR>70$ & -- & 1 & 1 & 0.8 & 0.25 & 0.93 & 0.76\\
    \hline
     nMR-precision& $AR<10$ & 1 & 0.74 & 0.71 & 0.65 & 0.90 & 0.7 & 0.75 \\
     & $AR<12.5$ & 1 & 0.76 & 0.74 & 0.63 & 0.90 & 0.71 & 0.76\\
     & $AR<20$ & 1 & 0.7 & 0.72 & 0.65 & 0.89 & 0.69 & 0.74\\
     & $AR<30$ & 1 & 0.66 & 0.68 & 0.64 & 0.88 & 0.66 & 0.71\\
     & $AR<50$ & 1 & 0.65 & 0.68 & 0.62 & 0.88 & 0.65 & 0.71\\
     & $AR<70$ & 1 & 0.61 & 0.63 & 0.60 & 0.87 & 0.61 & 0.68\\
    \hline
    \end{tabular}
    \caption{Number of structures found, precision and nMR-precision for A1, for different AR threshold (from 10 to 70). The results are shown for five different times of our simulation:  $t\sim20$ $[1/\Omega_{ci}]$, $t\sim 230$ $[1/\Omega_{ci}]$, $t\sim 247$ $[1/\Omega_{ci}]$, $t\sim 282$ $[1/\Omega_{ci}]$, $t\sim 494$ $[1/\Omega_{ci}]$. In the last two columns we report the mean value of our quality-parameters for the three central times (230, 247, 282) and for four times (230, 247, 282, and 494). In the second row we list the number of structures found at the end of the third step; in rows from 3rd to 8th we give the number of detected structures overcoming the specified AR threshold. In rows from 9th to 14th we give the values of precision for different AR threshold computed among the structures enumerated in rows 3-8; finally in rows 15th to 20th the same for nMR-precision.}
    \label{table:method_2}
    \end{table}

    
    \begin{table}[h!]
    \centering
    \begin{tabular}{c c|c c c c c | c c} 
    
     Tempo [1/$\Omega_{ci}$]& & 20 & 230 & 247 & 282 & 494 & Mean (3t) & Mean (4t)\\
    \hline
     N. structures & & 2 &  30 & 27 & 26 & 126  \\
     \hline
     N. structures &$AR>10$ & 0 & 21 & 18 & 12 & 52  \\
     &$AR>12.5$ & 0 & 19 & 17 & 11 & 43\\
     & $AR>20$ & 0 &  16 & 14 & 10 & 29\\
     & $AR>30$ & 0 &  13 & 11 & 9 & 20\\
     & $AR>50$ & 0 &  11 & 9 & 7 & 8\\
     & $AR>70$ & 0 &  6 & 6 & 5 & 3\\
     \hline
     precision & $AR>10$ & -- & 0.76 & 0.78 & 0.92 & 0.33 & 0.82 & 0.70\\
     & $AR>12.5$ & -- &  0.84 & 0.82 & 0.91 & 0.33 & 0.86 & 0.72\\
     & $AR>20$ & -- &  0.81 & 0.86 & 0.9 & 0.38& 0.86 & 0.74\\
     & $AR>30$ & -- & 0.77 & 0.91 & 0.89 & 0.4& 0.86 & 0.74\\
     & $AR>50$ & -- &  0.82 & 1 & 0.86 & 0.37 & 0.89 & 0.76\\
     & $AR>70$ & -- &  1 & 1 & 0.8 & 0.33 & 0.93 & 0.78\\
    \hline
     nMR-precision& $AR<10$ & 1  & 0.89 &  0.44 & 0.64 & 0.81 & 0.66 & 0.69\\
     & $AR<12.5$ & 1 & 0.91 &  0.5 & 0.6 & 0.79& 0.67 & 0.7\\
     & $AR<20$ & 1 & 0.71 &  0.46 & 0.56 & 0.79 & 0.58 & 0.63\\
     & $AR<30$ & 1 & 0.59 & 0.44 & 0.53 & 0.78 & 0.52 & 0.58\\
     & $AR<50$ & 1 & 0.58 & 0.44 & 0.47 &  0.77 & 0.5 & 0.56\\
     & $AR<70$ & 1 & 0.54 & 0.38 & 0.43 & 0.76  & 0.45 & 0.53\\
    \hline
    \end{tabular}
    \caption{Number of structures found, precision and nMR-precision for A2, for different AR threshold (from 10 to 70). The results are shown for five different times of our simulation:  $t\sim20$ $[1/\Omega_{ci}]$, $t\sim 230$ $[1/\Omega_{ci}]$, $t\sim 247$ $[1/\Omega_{ci}]$, $t\sim 282$ $[1/\Omega_{ci}]$, $t\sim 494$ $[1/\Omega_{ci}]$. In the last two columns we report the mean value of our quality-parameters for the three central times (230, 247, 282) and for four times (230, 247, 282, and 494). 
     In the second row we list the number of structures found at the end of the third step; in rows from 3rd to 8th we give the number of detected structures overcoming the specified AR threshold. In rows from 9th to 14th we give the values of precision for different AR threshold computed among the structures enumerated in rows 3-8; finally in rows 15th to 20th the same for nMR-precision.}
    \label{table:method_3}
    \end{table}
    
\section{Conclusions}
Summarizing, we developed three different Methods aimed at automatically detecting reconnecting CSs in 2D simulation of turbulence. AML uses unsupervised machine learning techniques for the scope, while Methods A1 and A2 use only thresholds on the standard reconnection proxies. All Methods are based on a threshold on the structures' AR since the AR is directly linked to the reconnection rate and is a physical adimensional parameter independent from all simulations set up. We have defined two different quality parameters in order to estimate quantitatively the performance of our algorithms: precision, which gives a measure of the capability of the method to select good reconnection sites, and nMR-precision, which gives us an information about excluded regions, in particular how many of them are really not reconnecting sites. In other words, it is an information about good sites excluded. AML is the one which performs the best. Moreover, although A2 has a better precision at smallest AR threshold, for AML nMR-precision is better, and it is possible to define for it an optimal AR (as the intersection between the plot of precision an nMR-precision) which instead cannot be done for A2. The impossibility to define an optimal AR for A2, together with the bad values of nMR-precision for this method (lots of good sites lost), brings us to the conclusion that A2 is not the best method to be applied. Considering only time periods where CSs are formed but still not interacting or significantly affected by the turbulent dynamics, we found for AML an optimal AR of about 18, which gives both a precision and a nMR-precision of $\sim 78\%$. 
These average performances worsen a bit if we consider also times at developed turbulence. This worsening is expected since at the fully-developed turbulent phase the dynamics advects, shrinks, breaks and deforms the CSs. Moreover, CS merging could lead to problems in separating different peaks and developed turbulence creates regions with sharp variations in current density where, however, reconnection is not occuring thus ``confusing'' the automatic Methods and decreasing their accuracy. Despite this worsening during the fully developed turbulent phase, the automatic Methods (and in particular the AML one) are still valid supports for speeding up reconnection site identification. In principle they could be applied to any plasma simulation including multiple potential reconnection sites, efficiently providing a list of ``candidate'' sites for a detailed human-based reconnection analysis. 

These automatic Methods are not based on the analysis of images as, for instance, 2D charts of simulation data, but use physical, measured variables as relevant signatures of reconnection. These quantites, here taken from numerical simulations, are also available in satellite data sets. We are presently working on adapting these methods to the analysis of data collected on 1D trajectories, more precisely cuts in the simulation plane or temporal series in the case of a satellite surveying plasma/magnetic structures. Such an upgrade would greatly improve the analysis of magnetic reconnection in satellite data, quickly highlighting interesting temporal intervals with active reconnection randomly scattered in the huge amount of data produced by past and present missions.

We stress also the importance of setting on a threshold for the CS's AR for better the performaces of our Methods. Indeed the aspect ratio of a reconnecting current sheet is a physical, adimensional, simulation independent parameter directly linked to the reconnection process. The increasing of the precision with the AR threshold value is linked to the fact that 
once the characteristic length of the reconnecting structure is fixed (by the boundaries of adjacent flux ropes),  reconnection is able to effectively shrink its width, leading to increasing values of aspect ratio. However, by
increasing the AR threshold too much above the theoretical estimation would lead to incorrect results since an excessive thinning corresponds to a strongly non linear phase eventually leading to
 to the break of the site in smaller magnetic islands. In fact nMR-precision worsens for too large AR threshold as reconnection itself limits the AR of the CS. As a consequence, we propose an optimal AR threshold for which both precision and nMR-precision perform well. We conclude by saying that the AML method, based on the machine learning approach, turns out to be the most performing one.\\

In the context of the EU AIDA project, we are presently working to set up a tool, namely Unsupervised ML Reconnection, in Python language for the AML method. This tool will become available, free-to-use, in the online AIDA-repository at \url{https://gitlab.com/aidaspace/aidapy} aimed at being of utility for the (space) plasma physics community. As for today, some preliminary scripts are available at \url{https://gitlab.com/aidaspace/aidapy/-/tree/unsupervised_reconnection/aidapy/WIP-unsupervised_reconnection}, and also on Zenodo at \url{https://doi.org/10.5281/zenodo.4282289}. Finally, the simulation data-set (TURB 2D) is available at Cineca AIDA-DB. In order to access the meta-information and the link to ``TURB 2D'' simulation data look at the tutorial at \url{http://aida-space.eu/AIDAdb-iRODS}.

\acknowledgements
This project has received funding from the European Union's Horizon 2020 research and innovation programme
under grant agreement No 776262 (AIDA, www.aida-space.eu).
Numerical simulations discussed here have been performed on Marconi at Cineca (Italy) under the ISCRA initiative. One of us, FC, thank Dr. M. Guarrasi (Cineca, Italy) for his essential contribution for the implementation of the kinetic Vlasov code on Marconi supercomputer at Cineca (Italy).







\bibliography{AAA_revised_version}{}
\bibliographystyle{aasjournal}



\end{document}